\documentclass[Physsubmission, Phys]{SciPost}
\hypersetup{
    colorlinks,
    linkcolor={red!50!black},
    citecolor={blue!50!black},
    urlcolor={blue!80!black}
}

\usepackage[bitstream-charter]{mathdesign}
\usepackage{xspace}
\DeclareSymbolFont{usualmathcal}{OMS}{cmsy}{m}{n}
\DeclareSymbolFontAlphabet{\mathcal}{usualmathcal}

\newcommand{\xmax}{\ensuremath{X_\mathrm{max}}\xspace}
\newcommand{\xmaxmed}{\ensuremath{\left< X_\mathrm{max} \right>}\xspace}
\newcommand{\sigXmax}{\ensuremath{\sigma(X_\mathrm{max})}\xspace}

\newcommand{\gcm}{\ensuremath{\mathrm{g~cm^{-2}}}\xspace}
\newcommand{\exprange}{\ensuremath{\mathrm {km^{2}~sr~yr}}\xspace}

\begin{document}

\begin{center}
\Large \textbf{
Astroparticle and particle physics at ultra-high energy:
results from the Pierre Auger Observatory}
\end{center}

% TODO: write the author list here. Use initials + surname format.
\begin{center}
Antonella Castellina\textsuperscript{$\star$},
for the Pierre Auger Collaboration\textsuperscript{$\star \star$}
\end{center}

\begin{center}
{\bf *} INFN, Sezione di Torino and Osservatorio Astrofisico di Torino (INAF), Torino, Italy \\
{\bf **} Full author list: \href{https://www.auger.org/archive/authors_2022_05.html}{http://www.auger.org/archive/authors\_2022\_05.html} \\
E-mail: spokespersons@auger.org
\end{center}

%\begin{center}
%$\star$ INFN, Sezione di Torino and Osservatorio Astrofisico di Torino (INAF), Torino, Italy \\
%$\star \star$ Full author list: 
%\href{mailto:auger_spokespersons@fnal.gov}
%\end{center}
%Observatorio Pierre Auger, Av.\ San Mart\'in Norte 304, 5613 Malarg\"ue, Argentina

%\begin{center}
%\today
%\end{center}
% For convenience during refereeing (optional),
% you can turn on line numbers by uncommenting the next line:
%\linenumbers

\definecolor{palegray}{gray}{0.95}
\begin{center}
\colorbox{palegray}{
  \begin{tabular}{rr}
  \begin{minipage}{0.1\textwidth}
    \includegraphics[width=30mm]{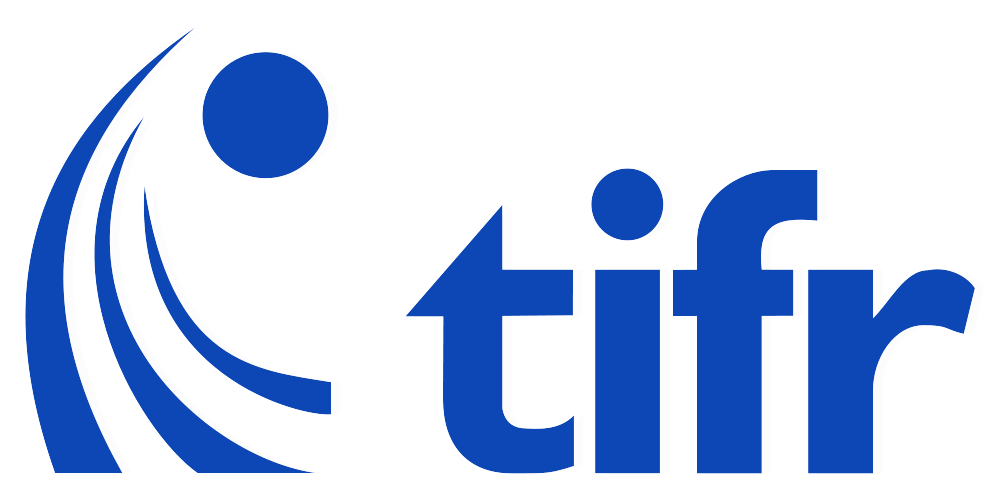}
  \end{minipage}
  &
  \begin{minipage}{0.85\textwidth}
    \begin{center}
    {\it 21st International Symposium on Very High Energy Cosmic Ray Interactions (ISVHE- CRI 2022)}\\
    {\it Online, 23-27 May 2022} \\
    \doi{10.21468/SciPostPhysProc.?}\\
    \end{center}
  \end{minipage}
\end{tabular}
}
\end{center}

\section*{Abstract}
{\bf
% TODO: write your abstract here.
%The abstract is in boldface, and should fit in 8 lines.
The scientific achievements of the Pierre Auger Collaboration cover diverse and complementary fields of research. The search for the origin of ultra-high energy cosmic rays (UHECRs) is based on the measurement of the energy spectrum and mass composition of the primaries, on studies of multi-messengers, 
%photons and neutrinos, 
and on extensive anisotropy searches. %at both large and intermediate angular scales. 
With the collected data it is also possible to explore the characteristics of hadronic interactions at energies unreachable at human-made accelerators, and to assess the existence of non-standard physics effects.
%, such as possible Lorentz invariance violations, or signals of super-heavy dark matter. 
A selection of the latest results is presented and the emerging picture is discussed.
%, starting from the quest for the sources of UHECRs to then focus on hadronically-sensitive shower observables.
%and their comparisons with model predictions. 
%Finally, the expected contribution of the ongoing upgrade AugerPrime of the Observatory will be discussed.
}

\vspace{10pt}
\noindent\rule{\textwidth}{1pt}
\tableofcontents\thispagestyle{fancy}
\noindent\rule{\textwidth}{1pt}
\vspace{10pt}

\section{Introduction}
\label{sec:intro}
The quest for the sources of ultra-high energy cosmic rays (UHECRs) is one of the most interesting open questions in astroparticle physics. Although we still lack a comprehensive theory capable of explaining where the UHECRs are produced, what is the acceleration mechanism driving nuclei to such extreme energies and their chemical composition, a large amount of information was gained in the recent years by exploiting the data taken with the Pierre Auger Observatory.
The largest sample of ultra-high energy events ever collected led to major advances in our understanding of their properties, reaching a global view which appears to be much more complex than the one available in the past. 
A key role in the interpretation of the results is played by the hadronic interaction models. Their predictions, based on the most up-to-date results by accelerators, need to be extrapolated to much higher energies. There, UHECRs are the only probes available; besides providing new information, they can also be exploited to explore possible effects beyond the standard model of particle physics.

The Pierre Auger Observatory \cite{nim} has collected data for more than 15 years, accumulating the world’s largest exposure to ultrahigh energy cosmic rays (UHECRs). Extensive air showers induced by CR-interactions in the atmosphere are observed using 24 fluorescence telescopes (FD) and a surface detector (SD1500) consisting of 1600 water Cherenkov stations distributed over an area of $3000$~km$^2$ at mutual distance of $1500$~m. Three more FD telescopes and a denser array of 61 detectors  
%at mutual distance of $750$~m 
(SD750) are deployed in the North-Western part of the Observatory, covering $27$~km$^2$, to extend the measurements to lower energies. Using the FD technique, we can measure the nearly calorimetric energy deposited in the atmosphere by the electromagnetic component of the EAS. 
The contribution of the so-called "invisible energy" (carried on by muons and neutrinos) is then added based on a data-driven method, to obtain the total energy of the primary.
The energy calibration is obtained exploiting the correlation of the SD energy estimator (the particle density at 1 km from the shower core, corrected by the attenuation) with the FD energy in hybrid events, i.e. those detected by both the SD and the FD. The calibration can then be applied to all SD events in an almost model independent way.
The FD energy resolution amounts to about 8\%, the overall systematic uncertainty on the energy scale to 14\% \cite{calen}.

\section{Experimental results}

\subsection{Arrival directions}

Searches for anisotropies in the arrival directions of UHECRs can be performed at lower energies on large angular scales (much larger than the angular resolution of the experiment). In this case, the cumulative flux from different sources could possibly be observed despite magnetic  deflections. 
A large-scale dipolar anisotropy in the arrival direction distribution of events recorded by the Pierre Auger Observatory was first observed in \cite{LSA2017}, and further explored in \cite{LSA2018,LSA2020}, under the assumption that higher multipoles are negligible. 
%The anisotropy, detected at more than $6.5 \sigma$ level of significance, can be described by a dipole with amplitude of 6:5þ1:3 percent
While for energies below $8 \times 10^{18}$~eV the result is consistent with isotropy, above this energy the first-harmonic modulation in right ascension leads to an equatorial dipole amplitude of $0.06 \pm 0.01$, with a probability to arise by chance from an isotropic distribution of $1.4 \times 10^{-9}$. 
%With more than 15 years of data, including events up to $80^{\circ}$, we could exploit for this analysis the enormous exposure of $120,000$ \exprange. 
The dipole points $\sim 125^{\circ}$ away from the Galactic centre direction, thus suggesting an extra-galactic nature of the UHECRs above this energy.  
A Galactic origin would result in a much stronger dipole, in a direction within a few tens of degrees of the Galactic Center. 
\begin{figure}[h]
\centering
\includegraphics[width=0.95\textwidth]{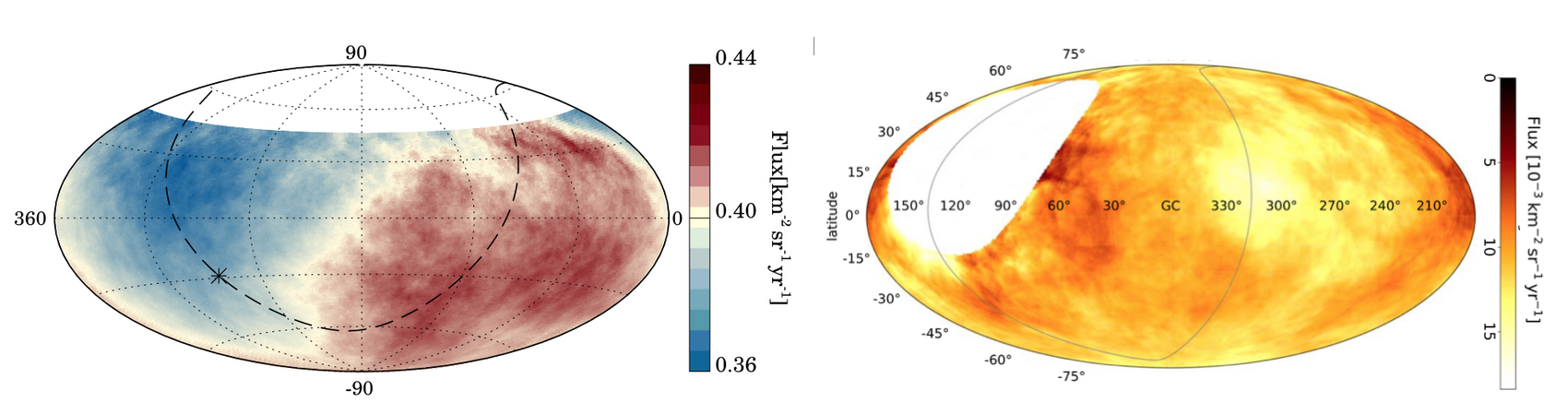}
\caption{Left: Flux map at energies above $8$~EeV, smoothed in windows of $45^{\circ}$. Right: Flux map at energies above $40$~EeV with a top-hat smoothing radius of $\Psi=25^{\circ}$.}
\label{fig:anis}
\end{figure}
%\vspace{-0.5cm}
%the upper bound to the amplitude being about $1.2 \%$ at $95 \%$ C.L.
%The statistical significance of this result is $6.6 \sigma$.\\
The dipole amplitude grows with energy; this could be associated with the larger relative contribution to the flux that arises at high energies from nearby sources.
A suppression of the more isotropic contribution from farther ones is expected to result from the strong attenuation of the CR flux that should take place at the highest energies as a consequence of their interactions with the background radiation. 
This energy dependence can be reproduced only with an evolution of the mass composition \cite{moll,ding}.

At higher energies, intermediate scale anisotropies can be searched for, as Galactic and extra-galactic deflections of UHECR are expected to be small enough for point sources to be visible as warm/hot spots.   
The largest available set of events at such energies was obtained thanks to an integrated exposure of $122,000$ \exprange. 
Evidence of a deviation from isotropy has been found by correlating the arrival directions of UHECRs with the direction of starburst galaxies and active galactic nuclei (AGN) above $\sim 40 \times 10^{18}$~eV, at $4.2 \sigma$ and $3.3 \sigma$ significance level respectively \cite{ISA2022}.
This result has been derived accounting for the attenuation of the UHECR 
nuclei, assuming a mixed mass composition as inferred from lower energy observations. 
The observed correlation does not allow us to disentangle different classes of objects yet; it could furthermore be altered by the effects of Galactic and extra-Galactic magnetic fields. 
The overall fluxes in the two energy bins are shown in Fig.\ref{fig:anis}.

A joint collaboration between the Pierre Auger Observatory and the Telescope Array (TA), a hybrid experiment operating since 2008 in the Northern hemisphere and covering an area of $700$~km$^2$, allows us to explore possible deviations from isotropy over the full sky.
The search for large scale anisotropies over the full sky confirmed the Auger results described above, with reduced systematic uncertainties 
%on the North-South components of the dipole and quadrupole components 
\cite{lsaPAOTA22}. 
Studies on intermediate scale anisotropies found excesses lying along the Supergalactic Plane; $11.8\%^{+5.0\%}_{-3.1\%}$ of the events show a  correlation with the position of nearby starburst galaxies on a $15^{\circ}$ angular scale, with a post-trial significance of $4.2 \sigma$ \cite{isaPAOTA22}. 

\subsection{Energy spectrum and composition}
\begin{figure}[h]
\centering
\includegraphics[width=0.95\textwidth]{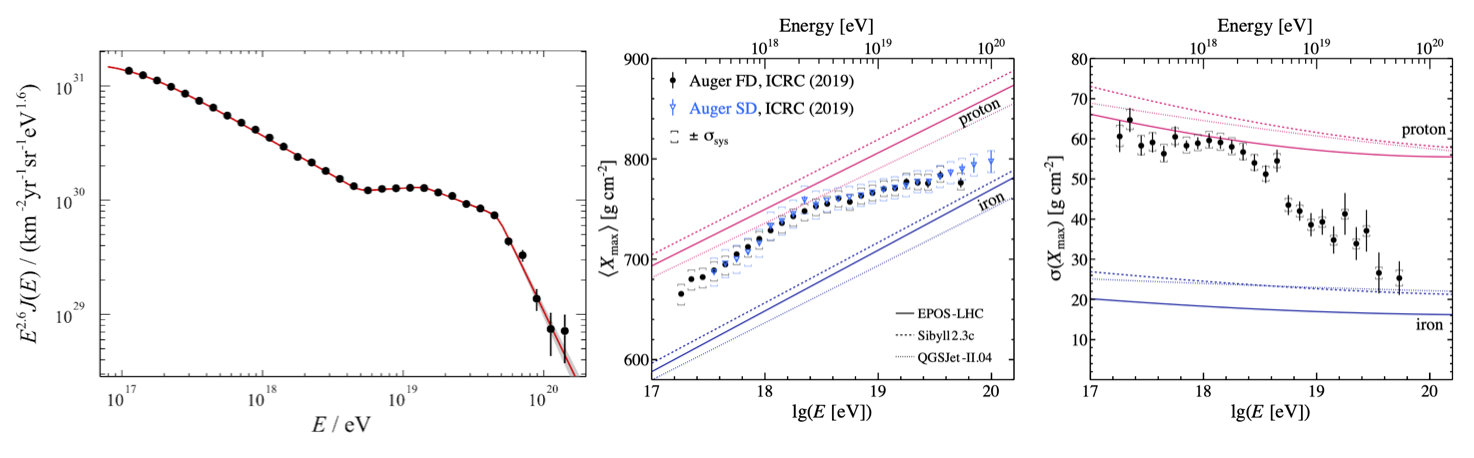}
\caption{Left: SD energy spectrum after combining the individual measurements by the SD-750 and the SD-1500 scaled by $E^{2.6}$. Right: the first two moments of the \xmax distributions measured by the FD.}
\label{fig:specomp}
\end{figure}

A huge exposure of about $80,000$ \exprange allowed us to measure the cosmic ray energy spectrum exploiting more than 800,000 events collected by the SD1500 and the SD750 arrays with energies above $10^{17}$~eV \cite{spec3}. 
The ankle and the flux suppression are clearly identified at $(4.9 \pm 0.1 \pm 0.8) \times 10^{18}$~eV and $(4.7 \pm 0.3 \pm 0.6) \times 10^{19}$~eV respectively, as shown in the left panel of Fig.\ref{fig:specomp}.
A new feature dubbed the {\it instep} could be identified at $1.4 \times 10^{19}$~eV.
The energy at which the integral spectrum drops by a factor of two below what would be that expected with no steepening is $E_{1/2}=(23 \pm 4) \times 10^{18}$~eV. This value is at variance with that expected for uniformly distributed sources of protons, as also indicated by the results from our composition measurements.

The measurement of the distributions of the depth of maximum development of the showers in the atmosphere provides us with the most reliable information on the UHECRs composition. 
Indeed, the average \xmaxmed is directly proportional to the logarithmic mass of the primary particle, while its variance is a convolution of the intrinsic shower-to-shower fluctuations and of the dispersion of masses in the primary beam.
The first two moments of the depth of shower maximum as measured at the Pierre Auger Observatory are shown in the central and rightmost panels of Fig.\ref{fig:specomp}. 
The observed rate of change of \xmaxmed with energy is $(77 \pm 2)$ and $(26 \pm 2)$ \gcm/decade below $E_{0}=10^{18.32\pm 0.03}$~eV and at higher energies respectively. 
The comparison with the predictions from models (tuned to the most recent LHC results) assuming pure primaries clearly points to a change of composition towards lighter elements up to $E_{0}$. Above this energy there is an increase in the average mass. 
Two data sets are plotted here: in black the direct measurement of \xmaxmed by the FD \cite{comp2014,comp2017}, in blue the SD result where the rise-time of the signal in the stations is used as the observable related to mass \cite{SDcomp}. 
This latter sample allows us to extend the composition measurements to higher energies, although at the expense of larger systematic uncertainties.
%with respect to the FD evaluation.
A confirmation of these results comes from the measurement of the standard deviation. 
The large values of \sigXmax below $E_{0}$ can be interpreted as coming from either light or mixed primaries, whereas the subsequent decrease would corresponds to a purer intermediate to heavy composition.
A completely independent measurement of the \xmax distributions was recently obtained exploiting the AERA radio array of the Pierre Auger Observatory, confirming the previous conclusion \cite{RDcomp}.
The average composition of UHECRs also appears to be heavier at low than at high Galactic latitudes, the cut being put after a pre-scan at Galactic latitute $|b|=30^{\circ}$ \cite{compAni22}. 
The post-trial significance of this result is currently $4.4 \sigma$. 
The interpretation of this result is however not straigthforward: e.g., the Galactic magnetic field could be responsible of the mass dependence of anisotropy, but the same effect could arise from the different horizons probed by different primaries.

It is interesting to compare our findings with those provided by TA.
The results on the flux of UHECRs are in agreement within the systematic uncertainties up to about $3 \times 10^{19}$~eV. The difference appearing at higher energies is still under discussion \cite{spePAOTA}. 
From the point of view of composition, a detailed comparison taking into account the detector effects shows that the TA result is compatible with the mixed composition measured by the Pierre Auger Observatory at least up to 10 EeV \cite{massPAOTA}. 
The lack of statistics does not allow TA to distinguish between a mixed, Auger-like composition and a proton one. 

\subsection{Multi-messengers}
With its high potential in the identification of photons and neutrinos, the Pierre Auger Observatory has a key role in the field of multi-messenger astrophysics. 
Astrophysical neutrinos and photons can be produced in the sources or in their environment, and cosmogenic ones are byproducts of the interactions of UHECRs with the radiation fields permeating the Universe. 
While photons can propagate along distances of the order of $\sim 4-5$~Mpc at $10^{18}$~eV, neutrinos allow us to probe sources up to cosmological distances. Their fluxes depend on the properties of the sources and on the composition of the primary beam.  
Photon showers are selected based on the fact that they present a more elongated profile in the atmosphere, and thus a larger \xmax, a steeper lateral distribution and a reduced production of muons with respect to hadronic showers. 
Neutrinos are tagged by selecting young horizontal showers with a large electromagnetic component close to the detector; these events can only be produced by neutrinos as in hadronic-induced cascades the electromagnetic component would be almost completely absorbed in the atmosphere and only muons could be detected in the SD. \begin{figure}[h]
\centering
\includegraphics[width=0.95\textwidth]{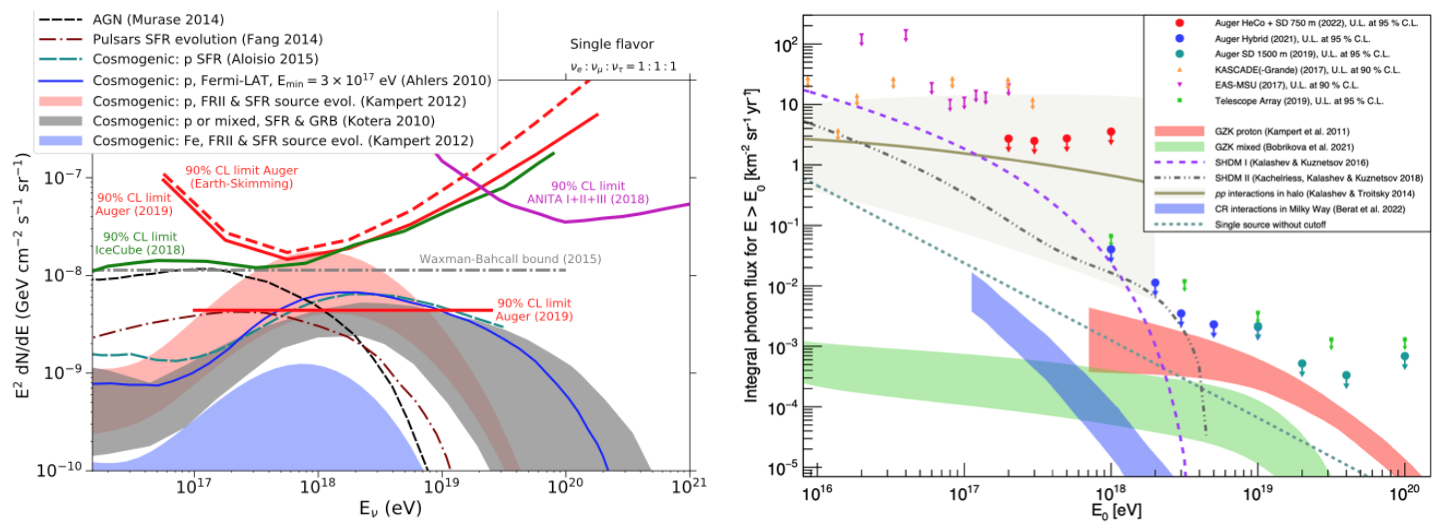}
\caption{Upper limits on the cosmogenic neutrinos (left) and photon (right) fluxes compared to limits from other experiments and to model predictions (\cite{neut2019,phot2022} and refs.therein).}
\label{fig:figmm}
\end{figure}
Two main categories of events are considered: Earth-skimming, induced by $\nu_{\tau}$ travelling from below the Earth crust in directions just below the horizon and producing a $\tau$ lepton which can then generate a shower above the SD, and downward-going events due to neutrinos of any flavour.
The most recent results on cosmogenic neutrinos and photons are shown in Fig.\ref{fig:figmm} \cite{neut2019,phot2022}.\\
Exploiting the very good angular resolution of the Observatory (less than $0.5^{\circ}$ for $\theta > 60^{\circ}$), it was also possible to look for point-like sources of UHE neutrinos \cite{neuPS,neuTXS}. 
Upper limits on the total energy per source were setup following up binary black hole mergers detected via gravitational waves (GW) \cite{mmNeu}; preliminary results on the photon fluence from selected GW sources were obtained in \cite{mmPho}.

\section{Constraining the sources of UHECRs}

\begin{figure}[h]
\centering
\includegraphics[width=0.95\textwidth]{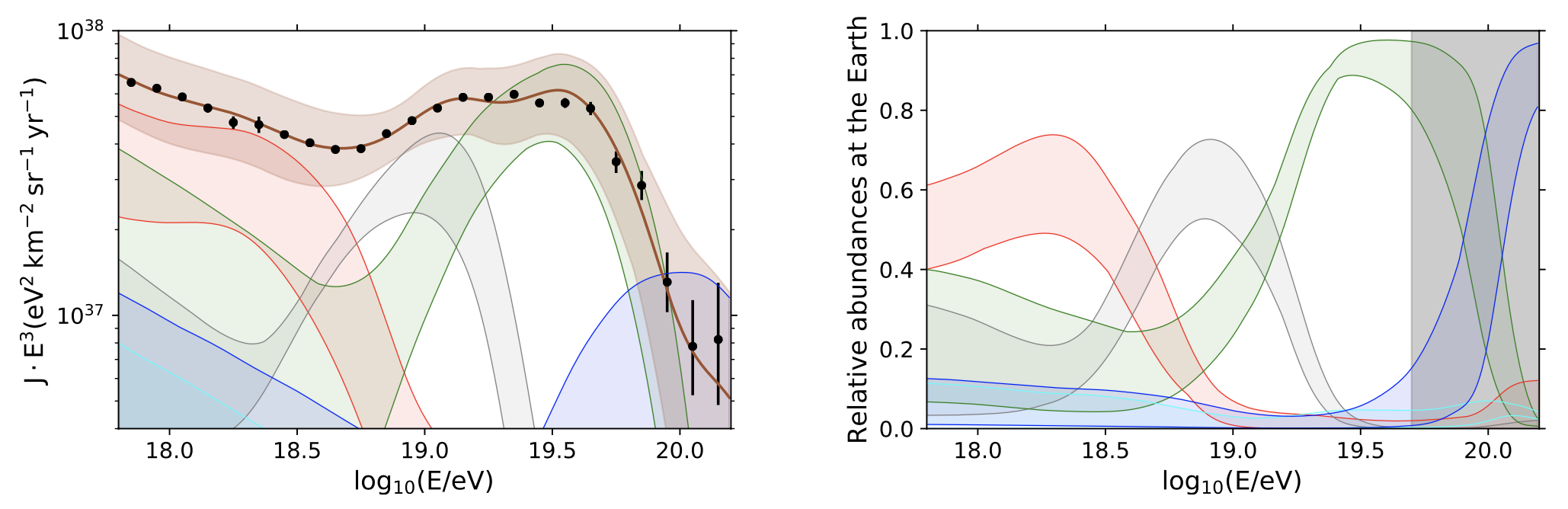}
\caption{Left: the energy spectrum. Right: the relative abundances at the top of atmosphere. The shaded areas indicate the combined effect of systematic uncertainties on $E$ and \xmax. 
[A=1 (red), 2 $\leq$ A $\leq$ 4 (grey), 5 $\leq$A$\leq$ 22 (green), 23 $\leq$A$\leq$ 38 (cyan), A $\geq$ 39 (blue)].}
\label{fig:ECF}
\end{figure}

Any astrophysical model, however simplified or complex, aiming at describing the UHECRs sources and their characteristics must be able to describe the particles features as measured at Earth. A comparison with the experimentally derived energy spectrum and composition is thus mandatory.\\
A combined fit of a simple astrophysical model of UHECR sources to both the energy spectrum and mass composition data measured by the Pierre Auger Observatory 
was exploited to investigate the constraining power of the collected data on the source properties, considering the energy region $E \gtrsim 6 \times 10^{17}$~eV) \cite{ECF2021}. 
Starting from a quite simple astrophysical scenario of stationary and uniform  sources in a co-moving volume, we considered several nuclear components injected at the sources with a power-law spectrum and with the maximal energy of the sources modeled with an exponential cut-off. \\
A good description of our data is obtained by considering two extragalactic components. 
Above the ankle, medium mass elements are escaping from the sources with a very hard energy spectrum and a rather low rigidity cutoff; the flux suppression appears to be mainly due to source exhaustion, rather than to propagation effects. 
The instep reflects the interplay between the flux contributions of the helium and the CNO group injected at the source, shaped by photo-disintegration during the propagation. \\
At lower energies, an additional light or light-to-intermediate very soft extragalactic component is needed. 
A similar result is obtained when this component is taken as an extragalactic proton-only one and a medium-mass Galactic contribution is added; this secondary Galactic component, if present, cannot be composed by heavy elements.\\
In Fig.\ref{fig:ECF}, the contribution of the different mass groups to the energy spectrum and the relative abundances at the top of the atmosphere are shown. 
The rather large uncertainty on the predicted total fluxes (brown band) is due to combined effects of the $\pm 14\%$ shifts in the energy scale and the uncertainty on \xmax, which ranges from 6 to 9 \gcm.
The experimental uncertainties are large, and mainly dominated by those affecting the measurement of \xmax; they must be taken into account in any models  comparing with data.
%As noted in \cite{spe2020b}, our data constrain the luminosity density that continuously emitting sources must inject into extragalactic space in UHECRs to supply the observed energy density to $\sim 6 \times 10^{44}$~erg Mpc$^{-3}$ yr$^{-1}$ above the ankle (for z=0).\\
As noted in \cite{spe2020b}, the luminosity density that continuously emitting sources must inject into extragalactic space in UHECRs 
to supply the observed energy density is constrained by our data  to $\sim 6 \times 10^{44}$~erg Mpc$^{-3}$ yr$^{-1}$ above the ankle (for z=0).
The unknown cosmological evolution of the sources strongly influences these results. Testing different combinations of source evolution for the two extragalactic components allows us to exclude very strong values (m=5) 
for the high energy one, as it would produce a too large flux of secondary particles at the ankle with respect to the measured one.

\section{UHECRs and particle physics}
%\label{sec:another}

\subsection{Hadronic interactions}
\begin{figure}[h]
\centering
\includegraphics[width=0.95\textwidth]{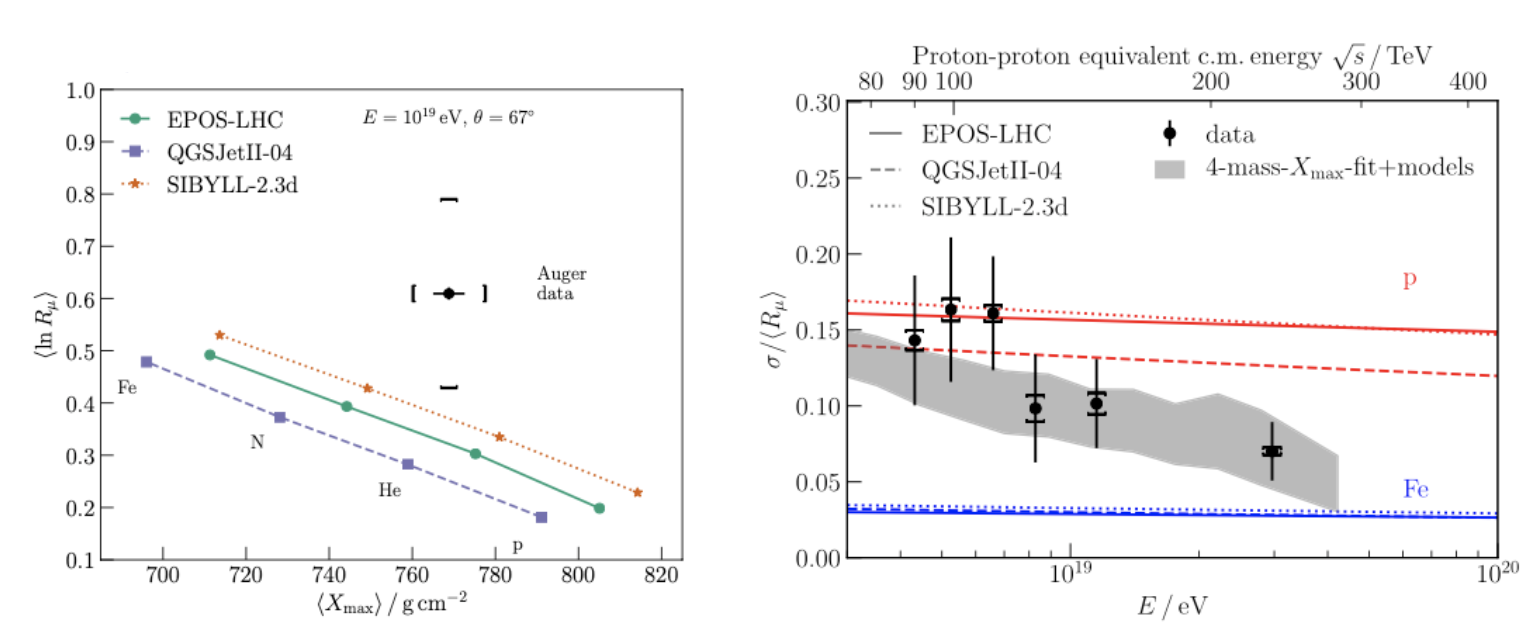}
\caption{Left: average logarithmic muon content $< ln R_{\mu} >$ as a function of the average shower depth \xmaxmed. 
Right: Measured relative fluctuations in the number of muons as a function of the energy. 
The grey band represents the expectations from the measured mass composition interpreted with the interaction models.}
\label{fig:mu}
\end{figure}

%%% copia di HIhglights, cambiare un po'
The interpretation of the experimental observables in terms of primary composition is prone to large systematic 
uncertainties, mainly due to the lack of knowledge on hadronic interactions at ultra-high energies.  \\
Measuring the attenuation length of proton-induced air showers (those mostly contributing to the tail of the \xmax distribution), we extended the measurement of the proton-air cross section up to 57~TeV \cite{pAir2012,pAir2015}. 
The measure of the muonic component at the ground is more sensitive to the details of the hadronic interactions along many steps of the cascade, like the 
multiplicity of the secondaries and the fraction of electromagnetic component with respect to the total signal. On the contrary, the intrinsic muon fluctuations mostly depend on the first interaction.\\
Exploiting inclined showers detected by the SD1500, where the electromagnetic component has been fully absorbed by the atmosphere \cite{muhor2015} (left panel of Fig.\ref{fig:mu}), or  hybrid events \cite{muhad2017}, we reported clear evidence for a deficit in the number of muons predicted by the current hadronic interaction models.
%Clear evidence for a deficit in the number of muons predicted by the models was reported by our Collaboration by exploiting inclined showers detected by the SD1500, where the electromagnetic component has been fully absorbed by the atmosphere \cite{muhor2015} (see left panel of Fig.\ref{fig:mu}) and  hybrid events \cite{muhad2017}. 
Direct measurements at lower energies confirmed the result \cite{mudir2020}. 
The problems in the modelling of hadronic interactions at ultra-high energy have further been proven by 
studying the muon production depth \cite{mudep2014} and the  time profiles of the signals recorded with SD \cite{delta}. \\
The measured shower-to-shower fluctuations of the muonic component are shown in the right panel of Fig.\ref{fig:mu} \cite{mufl2021}; 
the expectations for their relative values are compatible with the experimental results, while a significant discrepancy between models and data appears in the average muon scale. 
Note that  systematic errors (square brackets) are dominant for $\left< R_{\mu} \right>$, while for the muon fluctuations the uncertainties are mainly statistical (error bars).\\
A new method to explore simple ad-hoc adjustments to the predictions of hadronic interaction models was recently developed,
exploiting the correlation of \xmax with the signal at ground level as a function of the zenith angle \cite{vicha2021}. 
The uncertainties on the \xmax scale among different hadronic interaction models are of the order of 30 \gcm; on the other hand, the evaluation of the primary mass based on the signals at ground are prone to the inability of models to describe the muonic component. 
This method allows us to use both the longitudinal and lateral development of air showers to evaluate the flawns in the models. 
An overall improvement in the ability of simulations to describe the data is reached by shifting the expected depths of shower maximum deeper in atmosphere. 
At the same time, the hadronic signal at 1000 m from the core must be increased by a factor from 15 to 25\%, smaller than the one obtained with the previous methods.
This preliminary result will be further investigated by considering less simple adjustments to predictions, by taking into account also the fluctuations in both \xmax and signal at ground.

\subsection{Beyond the standard model}
Various quantum gravity theories suggest that the Lorentz invariance may be violated when approaching the Planck scale, resulting in a perturbative 
modification of the particle dispersion relation  
$E_{i}^{2}=p_{i}^{2}+m_{i}^{2}+\sum_{n=0}^{\infty} \delta^{(n)}_{i}E_{i}^{n+2}$
(where $i$=type of particle, $n$=approximation order). \\
The UHECRs collected at the Pierre Auger Observatory, coming from extragalactic sources, 
represent the ideal candidates to search for these effects, as these are expected to be suppressed at 
low energies and for short travel distances. 
In presence of Lorentz invariance violation (LIV), the attenuation length of photo-meson production or photo-disintegration may become extremely large and suppress particle interaction during propagation in the extragalactic space. 
Introducing these modifications in either the hadron or the photon propagation,
constraints on the level of violation can be set by comparing via a best fit procedure the obtained flux and composition to those measured at the top of the atmosphere \cite{LIV2022}. 
While in the case of the hadron sector the sensitivity to possible violations is very low, in the photon sector we obtained constraining limits of 
$\delta^{(1)}_{\gamma} \sim 10^{-40}$ eV$^{-1}$ and $\delta^{(2)}_{\gamma} \sim 10^{-60}$ eV$^{-1}$ for 
$n$=1 and $n$=2 respectively. \\
Depending on the strength of the violation, the high energy available in the collision of cosmic rays with the atmosphere can futhermore lead to modifications of the shower development with respect to the standard invariance case. Indeed, due to the much increased probability of interaction for neutral pions, the hadronic component of the cascade is  enhanced, leading in particular to a larger production of muons.
In this case, limits on the LIV parameter can be derived comparing the observed strong decrease of the relative fluctuations with the muon fluctuation measurement\cite{trim2022}.

Super-heavy relics produced in the early Universe could decay into UHE photons; detecting a flux of astrophysical photons with energies in excess of $\sim 10^{17}$~eV would thus be a proof of the existence of super-heavy dark matter. 
The non observation of these photons in the Pierre Auger Observatory allowed us to derive a bound on the reduced coupling constant of gauge interactions in the dark sector: $\alpha_{eff} \lesssim 0.09$ for $10^{10} < M_{X}/GeV < 10^{16}$ (\cite{delignyThis} and refs. therein).

\section{Conclusion}
The Pierre Auger Observatory is providing the best and most precise data world-wide on ultra-high energy cosmic rays, and a global view emerges from the analyses of collected data.
The suppression of the UHECR flux, measured with extremely high statistical significance, appears to be due to source exhaustion, although a contribution from propagation effects must be also present. 
This result is supported by the increasingly heavy nuclear composition in this energy region, and by the non observation of cosmogenic neutrinos and photons, a copious production of which should be expected for a proton dominated composition. 
The transition from Galactic to extra-galactic cosmic rays can be placed at the second knee. 
This is supported by the measured composition, which becomes lighter above the 2nd knee up to $\sim 2 \times 10^{18}$~eV, and by the smooth transition from isotropy
to a dipolar anisotropy above 8 EeV, where the phase of the dipole points to the anti-center of our Galaxy. 
Based on the results of the combined fit extended to the region below the ankle, two extragalactic populations, the low energy one made by proton/light elements, the high energy one by a mixture of nuclei (up to iron), can describe well the transition region. 
If the extragalactic component below the ankle is composed only by protons, a secondary Galactic component of intermediate nuclei (CNO) must be present.
The new instep feature can be explained as due to the interplay of He and CNO nuclei.
The presence of heavy nuclei at the highest energies implies a large effect of magnetic fields, the study of which becomes more and more mandatory. 
It is still to be clarified if the onset of the different components is due to propagation effects ($\propto A$) or to Peters cycle ($\propto Z$).  
Correlations have been found between the arrival directions of UHECRs and the direction of starburst Galaxies and AGN. About 3-4 years more data are needed to reach the $5 \sigma$ significance to claim observation. 
Disentangling the two types of sources remains an issue, as some of them (e.g. NGC4945 or NGC1068) are indeed composite AGN/SBG objects.
The strong limits on cosmogenic neutrinos and photons constrain models predicting proton sources; top-down scenarios are excluded. 
The deficit in the muon component predicted by models with respect to data is confirmed by many different analyses of independent data sets. 
The recent measurement of the fluctuations in the muonic component show that models describe reasonably well the energy partition in the first interaction up to the highest energies, 
while the discrepancy in the overall muon number should be explained by changes of different hadronic interaction characteristics along all the shower stages.
The data collected by the Pierre Auger Observatory allow us to explore the high energy frontier even beyond the standard model of particles. 
Constraining limits to a possible violation of the Lorentz invariance and on the mass and lifetime of super-heavy dark matter particles have been obtained. \\
New data on composition in the suppression region and more observables, like those that will be provided by the ongoing upgrade of the observatory, AugerPrime \cite{Prime2019}, will allow us to face the challenges these findings open to us for the future.

\footnotesize{

}
\nolinenumbers
\end{document}